  \documentstyle[11pt,menu97,epsfig]{article}

\newcommand{\be}{\begin{eqnarray}}
\newcommand{\ba}{\begin{array}}
\newcommand{\ea}{\end{array}}
\newcommand{\ee}{\end{eqnarray}}
\newcommand{\dslash}{\partial \hskip -0.5em /}

\begin{document}
    \setlength{\baselineskip}{3.2ex}
\rightline{UNITU-THEP-16/1997}
\rightline{OKHEP-97-03}
\rightline{July 1997}
\bigskip
\title{\Large \bf Nucleon Structure Functions within a Chiral Soliton 
Model\thanks{Talk presented by LG at the MENU97 Conference, Vancouver
July $28^{\rm th}$--Aug. $2^{\rm nd}$ 1997.
Work supported in part by the Deutsche 
Forschungsgemeinschaft (DFG) under contract Re 856/2--3 and the 
US--DOE grant DE--FE--02--95ER40923.}}
\author{L. Gamberg$^{a)}$, H. Reinhardt$^{b)}$ and H. Weigel$^{b)}$\\
{\em $^{a)}$Department of Physics and Astronomy, 
University of Oklahoma\\
440 W. Brooks Ave, Norman, Oklahoma 73019--0225, USA\\
$^{b)}$Institute for Theoretical Physics, T\"ubingen University\\
Auf der Morgenstelle 14, D-72076 T\"ubingen, Germany}}
\maketitle

\begin{abstract}
\setlength{\baselineskip}{3.2ex}
We study nucleon structure functions within the bosonized
Nambu--Jona--Lasinio model where the nucleon emerges as a
chiral soliton.  We discuss the model predictions on the Gottfried
sum rule for electron--nucleon scattering. A comparison with a
low--scale parametrization shows that the model reproduces the gross
features of the empirical structure functions. We also compute the
leading twist contributions of the polarized structure functions $g_1$
and $g_2$ in this model. We compare the model predictions on these
structure functions with data from the E143 experiment by GLAP evolving
them appropriately.
\end{abstract}

\setlength{\baselineskip}{3.2ex}

Here we discuss a possible link between two successful although 
seemingly unrelated pictures of baryons. On one side the quark parton 
model successfully describes the scaling behavior of the 
structure functions in deep inelastic scattering (DIS) processes. 
Scaling violations are computed within perturbative QCD. On the other 
side the chiral soliton approach, motivated by the large $N_C$ (number 
of colors) expansion of QCD, gives a fair account of many static 
baryon properties. For $N_C\to\infty$, QCD is equivalent to an 
effective theory of weakly interacting mesons. 
This unknown meson theory is modeled by rebuilding 
the symmetry structure of QCD. In particular this concerns chiral 
symmetry and its spontaneous breaking. Baryons then emerge as 
non--perturbative (topological) meson fields configurations, the 
so--called solitons. It is our goal to link these two pictures by 
computing nucleon structure functions within a chiral soliton model. We
start from the hadronic tensor  
\be
W^{ab}_{\mu\nu}(q)=\frac{1}{4\pi}\int d^4 \xi \
{\rm e}^{iq\cdot\xi}
\langle N(P) |\left[J^a_\mu(\xi),
J^{b{\dag}}_\nu(0)\right]|N(P)\rangle \ ,
\label{deften}
\ee
which describes the strong interaction part of the DIS 
cross--section. $|N(P)\rangle$ refers to the nucleon state with 
momentum $P$ and $J^a_\mu(\xi)$ to the hadronic current. The 
leading twist contribution to the structure functions is extracted 
from $W^{ab}_{\mu\nu}(q)$ by assuming the Bjorken limit
\be
q_0=|\mbox{\boldmath $q$}| - M_N x
\quad {\rm with}\quad
|\mbox{\boldmath $q$}|\rightarrow \infty
\quad {\rm and}\quad
x={-q^2}/{2P\cdot q}\quad  {\rm fixed}\ .
\label{bjlimit}
\ee

In most soliton models the current commutator (\ref{deften}) is
intractable. As an exception the Nambu and Jona--Lasinio 
(NJL) model \cite{Na61} for the quark flavor dynamics, which can be 
bosonized by functional integral techniques \cite{Eb86}, contains 
simple current operators. Most importantly, its bosonized version 
contains soliton solutions \cite{Al96}. We confine ourselves to 
the key issues of the structure function calculation as details 
are given in refs. \cite{We96,We97a,We97b}.

\section*{The Nucleon from the Chiral Soliton in the NJL Model}

Here we briefly summarize the description of baryons as 
chiral solitons in the NJL--model, details are found in 
the review articles \cite{Al96,Ch96}. We consider a chirally
symmetric NJL model Lagrangian which contains interactions 
in the pseudoscalar channel. Derivatives of the quarks fields 
only appear in form of a free Dirac Lagrangian, hence the current 
operator is formally free. Upon bosonization the action may 
be expressed as \cite{Eb86}
\be
{\cal A}={\rm Tr} \ {\rm ln}_\Lambda
\left(i\dslash - m U^{\gamma_5}\right)
+\frac{m_0m}{4G}{\rm tr}\left(U+U^{\dag}-2\right) \ ,
\label{bosact}
\ee
where ${\rm tr}$ and ${\rm Tr}$
denote discrete flavor and functional traces, respectively. 
The model parameters are the coupling 
constant $G$, the current quark mass $m_0$ and the UV cut--off 
$\Lambda$. The constituent quark mass $m$ arises as the solution to 
the Schwinger--Dyson (gap) equation and characterizes 
the spontaneous breaking of chiral symmetry. A Bethe--Salpeter 
equation for the pion field 
($U={\rm exp}(i\mbox{\boldmath $\tau$}\cdot
\mbox{\boldmath $\pi$}/f_\pi)$) is derived from eq. (\ref{bosact}).
Then the model parameters are functions of the pion mass 
$m_\pi=135{\rm MeV}$ and decay constant $f_\pi=93{\rm MeV}$. 
The sole undetermined parameter is $m$.
An energy functional for non--perturbative but
static field configurations $U(\mbox{\boldmath $r$})$ is 
extracted from (\ref{bosact}). It is expressed as a 
regularized sum of all single quark energies $\epsilon_\mu$. 
For the hedgehog {\it ansatz}, 
$U_H={\rm exp}(i\mbox{\boldmath $\tau$}\cdot
{\hat{\mbox{\boldmath $r$}}}\Theta(r))$ 
the one--particle Dirac Hamiltonian reads
\be
h=\mbox{\boldmath $\alpha$}\cdot\mbox{\boldmath $p$}
-m\ {\rm exp}\left(i\gamma_5\mbox{\boldmath $\tau$}\cdot
{\hat{\mbox{\boldmath $r$}}}\Theta(r)\right)\ , \quad
h\Psi_\mu = \epsilon_\mu\Psi_\mu \ .
\label{dirham}
\ee
The distinct level (v), bound in the background of $U_H$, is 
so--called valence quark state. Its explicit occupation 
guarantees unit baryon number. The self--consistent minimization 
of the energy functional determines the chiral angle $\Theta(r)$.
The so--constructed soliton does not carry nucleon quantum numbers. To 
generate them, the time dependent field configuration is approximated 
by elevating the zero modes to 
time dependent collective coordinates $A(t)\in {\rm SU}(2):$
$U(\mbox{\boldmath $r$},t)=A(t)U_H(\mbox{\boldmath $r$})A^{\dag}(t)$. 
Canonical quantization of the angular velocities,
$\mbox{\boldmath $\Omega$}=-2i{\rm tr}
(\mbox{\boldmath $\tau$}A^{\dag}\dot A)$, introduces the 
spin operator $\mbox{\boldmath $J$}$ via 
$\mbox{\boldmath $\Omega$}=\mbox{\boldmath $J$}/\alpha^2$
with $\alpha^2$ being the moment of inertia\footnote{Generalizing this
treatment to flavor SU(3) indeed shows that the baryons have to be 
quantized as half--integer objects. For a review on solitons in 
SU(3) see {\it e.g.} \cite{We96a}.}. The nucleon states
$|N\rangle$ emerge as Wigner $D$--functions. The action (\ref{bosact}) 
is expanded in powers of $\mbox{\boldmath $\Omega$}$ corresponding 
to an expansion in $1/N_C$. In particular the valence quark 
wave--function $\Psi_{\rm v}(\mbox{\boldmath $x$})$ acquires a 
linear correction 
\be
\Psi_{\rm v}(\mbox{\boldmath $x$},t)=
{\rm e}^{-i\epsilon_{\rm v}t}A(t)
\left\{\Psi_{\rm v}(\mbox{\boldmath $x$})
+\sum_{\mu\ne{\rm v}}
\Psi_\mu(\mbox{\boldmath $x$})
\frac{\langle \mu |\mbox{\boldmath $\tau$}\cdot
\mbox{\boldmath $\Omega$}|{\rm v}\rangle}
{2(\epsilon_{\rm v}-\epsilon_\mu)}\right\}=
{\rm e}^{-i\epsilon_{\rm v}t}A(t)
\psi_{\rm v}(\mbox{\boldmath $x$}).
\label{valrot}
\ee
Here $\psi_{\rm v}(\mbox{\boldmath $x$})$ denotes the spatial part
of the body--fixed valence quark wave--function with the rotational
corrections included.

\section*{Structure Functions in the Valence Quark Approximation}

The starting point for computing the unpolarized structure
functions is the symmetric part of the hadronic tensor in a form 
suitable for localized quark fields \cite{Ja75}, 
\be
W^{lm}_{\{\mu\nu\}}(q)&=&\zeta\int \frac{d^4k}{(2\pi)^4} \
S_{\mu\rho\nu\sigma}\ k^\rho\
{\rm sgn}\left(k_0\right) \ \delta\left(k^2\right)
\int_{-\infty}^{+\infty} dt \int d^3x_1 \ d^3x_2 \
{\rm e}^{i(k_0+q_0)t}
\label{stpnt} \\* && \hspace{-2.1cm}
\times {\rm exp}\left[-i(\mbox{\boldmath $k$}+\mbox{\boldmath $q$})\cdot
(\mbox{\boldmath $x$}_1-\mbox{\boldmath $x$}_2)\right]
%\nonumber \\ && \hspace{0.1cm}
\langle N |\left\{
{\hat{\bar \Psi}}(\mbox{\boldmath $x$}_1,t)t_l t_m\gamma^\sigma
{\hat\Psi}(\mbox{\boldmath $x$}_2,0)-
{\hat {\bar \Psi}}(\mbox{\boldmath $x$}_2,0)t_m t_l\gamma^\sigma
{\hat\Psi}(\mbox{\boldmath $x$}_1,t)\right\}| N \rangle .
\nonumber
\ee
Note that the quark spinors are functionals of the soliton.
Here $S_{\mu\rho\nu\sigma}=g_{\mu\rho}g_{\nu\sigma}
+g_{\mu\sigma}g_{\nu\rho}-g_{\mu\nu}g_{\rho\sigma}$ and
$\zeta=1(2)$ for the structure functions associated with the
vector (weak) current and $t_m$ is a suitable isospin matrix. 
The matrix element between the nucleon states ($|N\rangle$) is 
taken in the space of the collective coordinates. Eq. (\ref{stpnt}) 
is derived by assuming the {\it free} correlation function 
for the intermediate quark fields. In the limit (\ref{bjlimit}) 
the momentum, $k$, of the intermediate quark is highly off--shell 
and hence not sensitive to momenta typical for the soliton configuration. 
Thus the use of the free correlation function is a valid treatment.

\begin{figure}
\centerline{
\hspace{-0.5cm}
\epsfig{figure=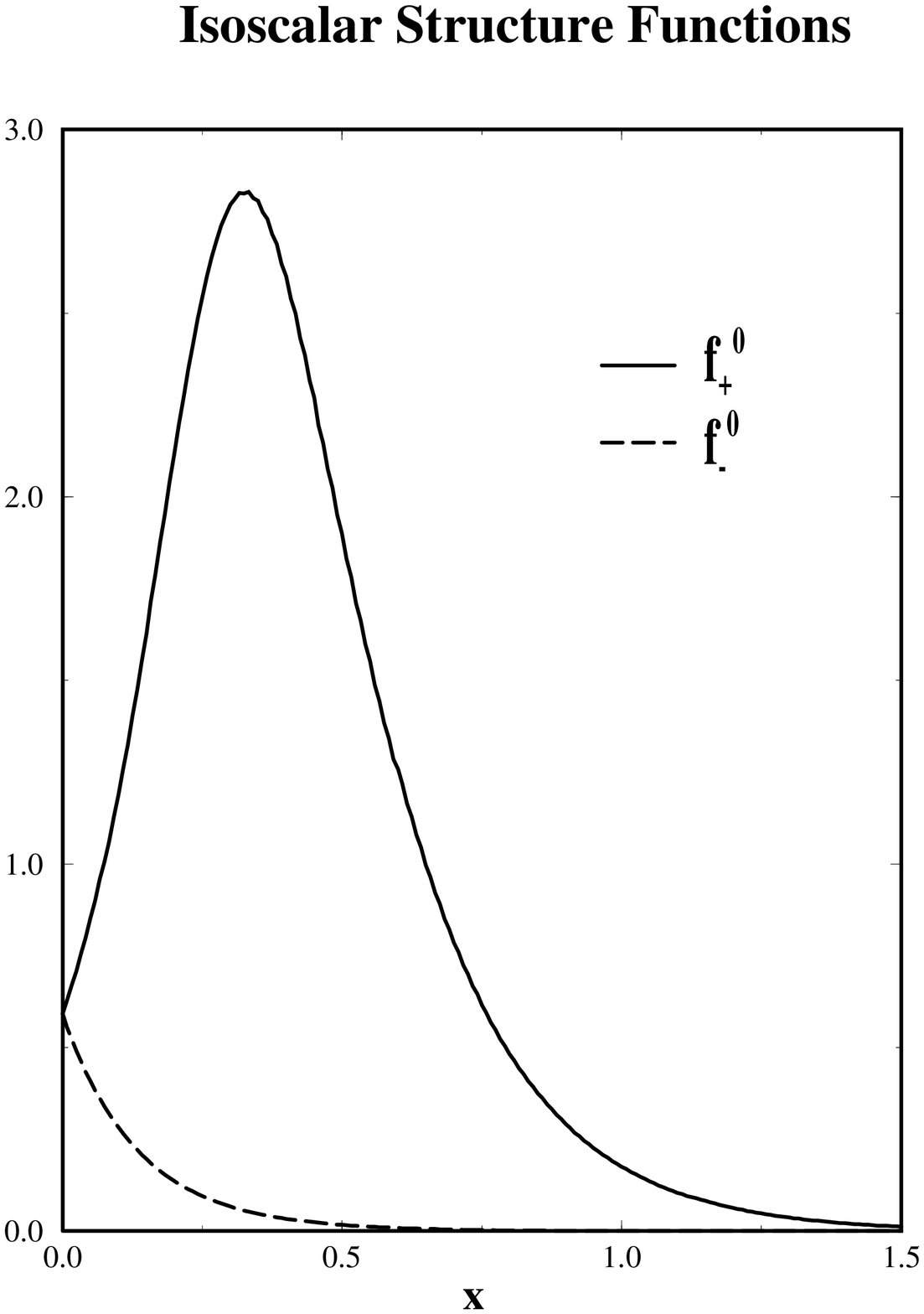,height=5.5cm,width=7.0cm}
\hspace{0.5cm}
\epsfig{figure=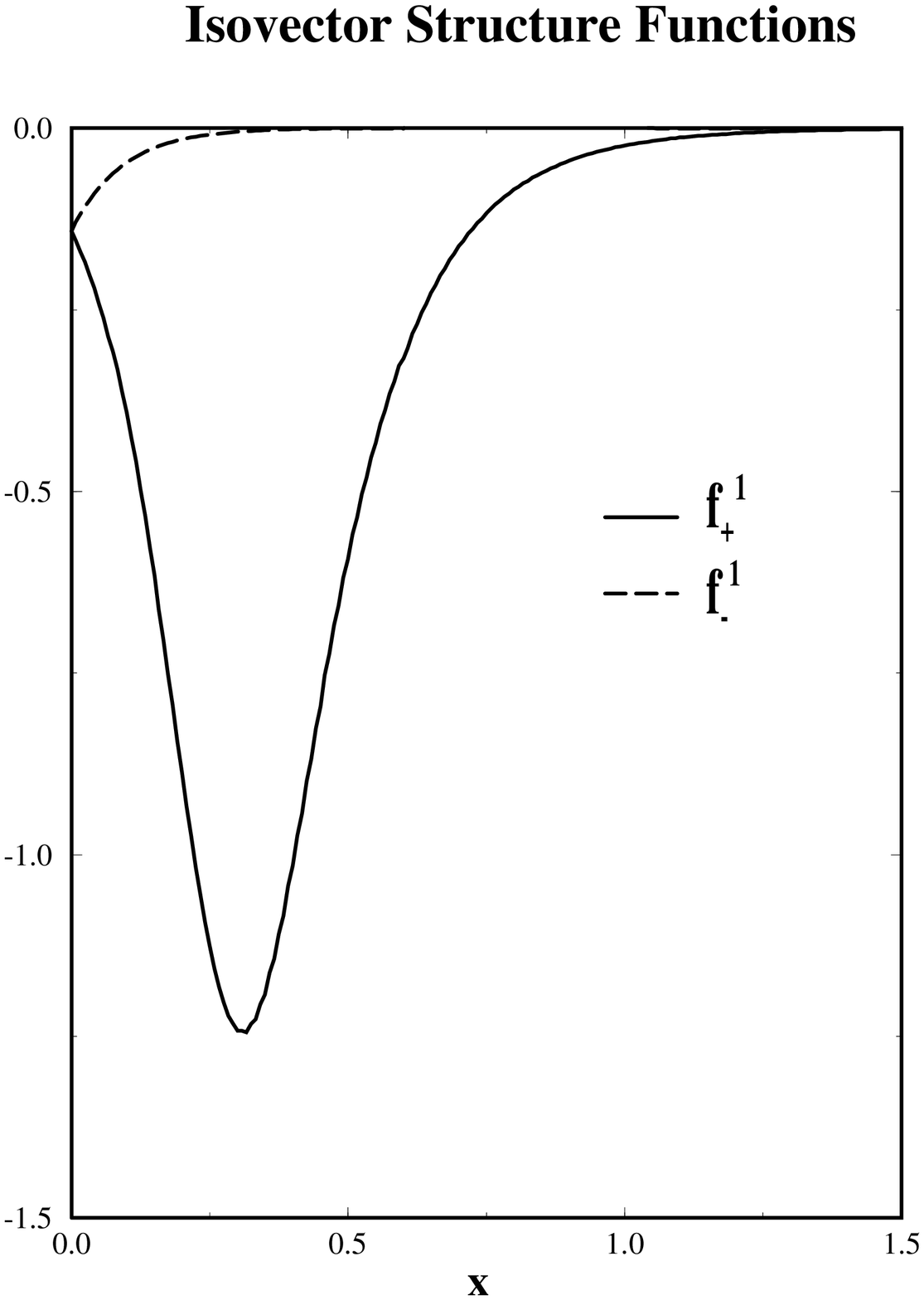,height=5.5cm,width=7.0cm}}
\vspace{-0.1cm}
\caption{\label{fig_1}The unpolarized structure functions
obtained after extracting the collective part of the
nucleon matrix elements. Here we used $m=350{\rm MeV}$.}
\vspace{-.3cm}
\end{figure}

The valence quark approximation ignores the vacuum polarization in 
(\ref{stpnt}), {\it e.g.} the quark field operator ${\hat\Psi}$ is 
substituted by the valence quark contribution (\ref{valrot}). For 
$m\sim400{\rm MeV}$ this is well justified since this level dominates 
the nucleon observables \cite{Al96,Ch96}. The structure function 
$F_2(x)$ is obtained from (\ref{stpnt}) by an appropriate 
projection\footnote{In the Bjorken limit the Callan--Gross relation 
$F_2(x)=2x F_1(x)$ is satisfied.}. After computing the collective 
coordinate matrix elements all physical relevant processes are 
described in terms of four reduced structure functions $f_\pm^{0,1}$
shown in figure \ref{fig_1}. The superscript denotes the isospin 
of $t_l\times t_m$ while the subscript refers to forward and 
backward moving intermediate quarks in (\ref{stpnt}).  
Although the problem is not formulated Lorentz--covariantly these 
structure functions are reasonably well localized in the interval 
$x\in [0,1]$. The contributions of the backward moving 
quarks are quite small, however, they increase with $m$. 

\begin{figure}[t]
\centerline{
\epsfig{figure=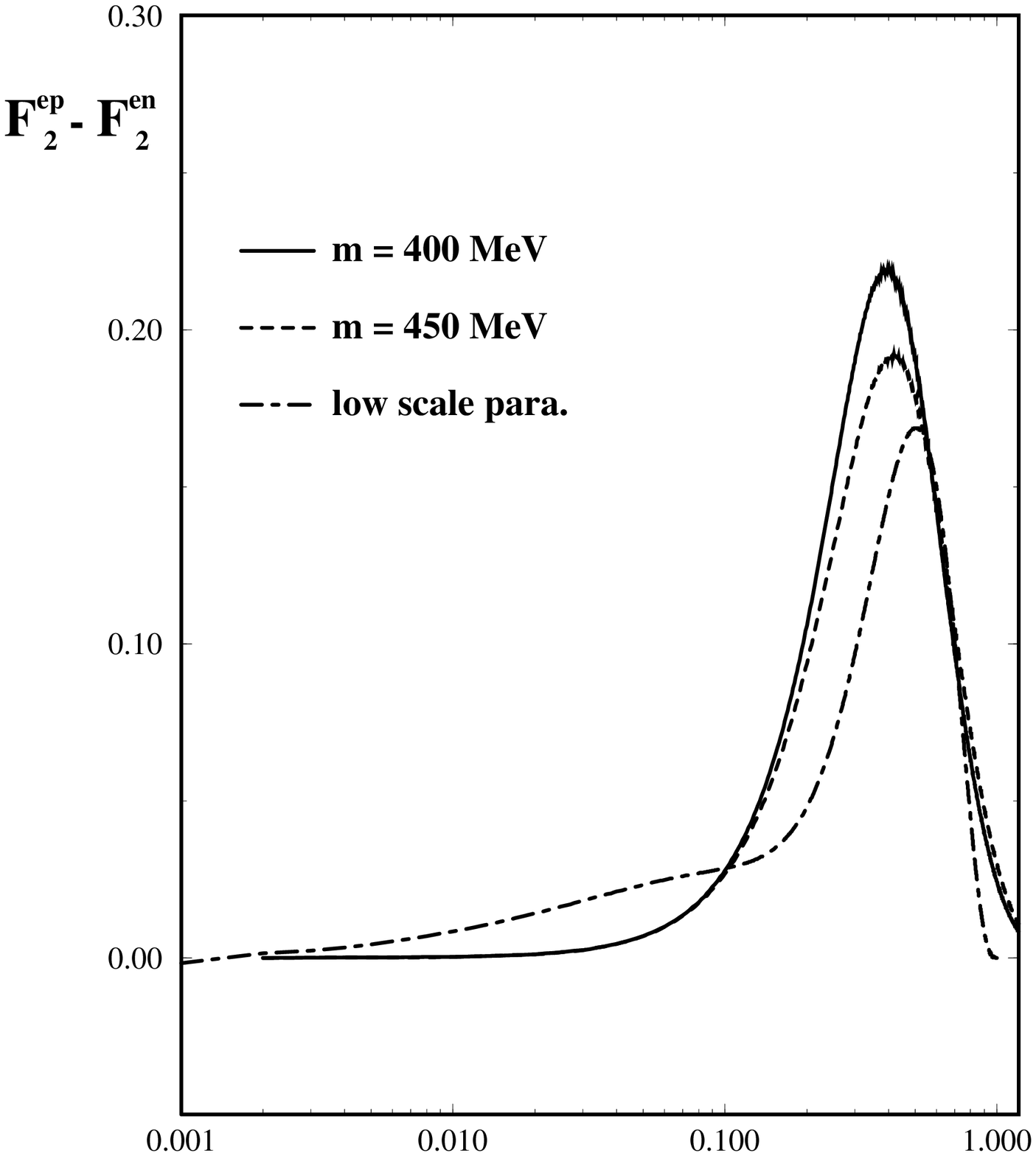,height=5.6cm,width=7.0cm}
\hspace{0.5cm}
\epsfig{figure=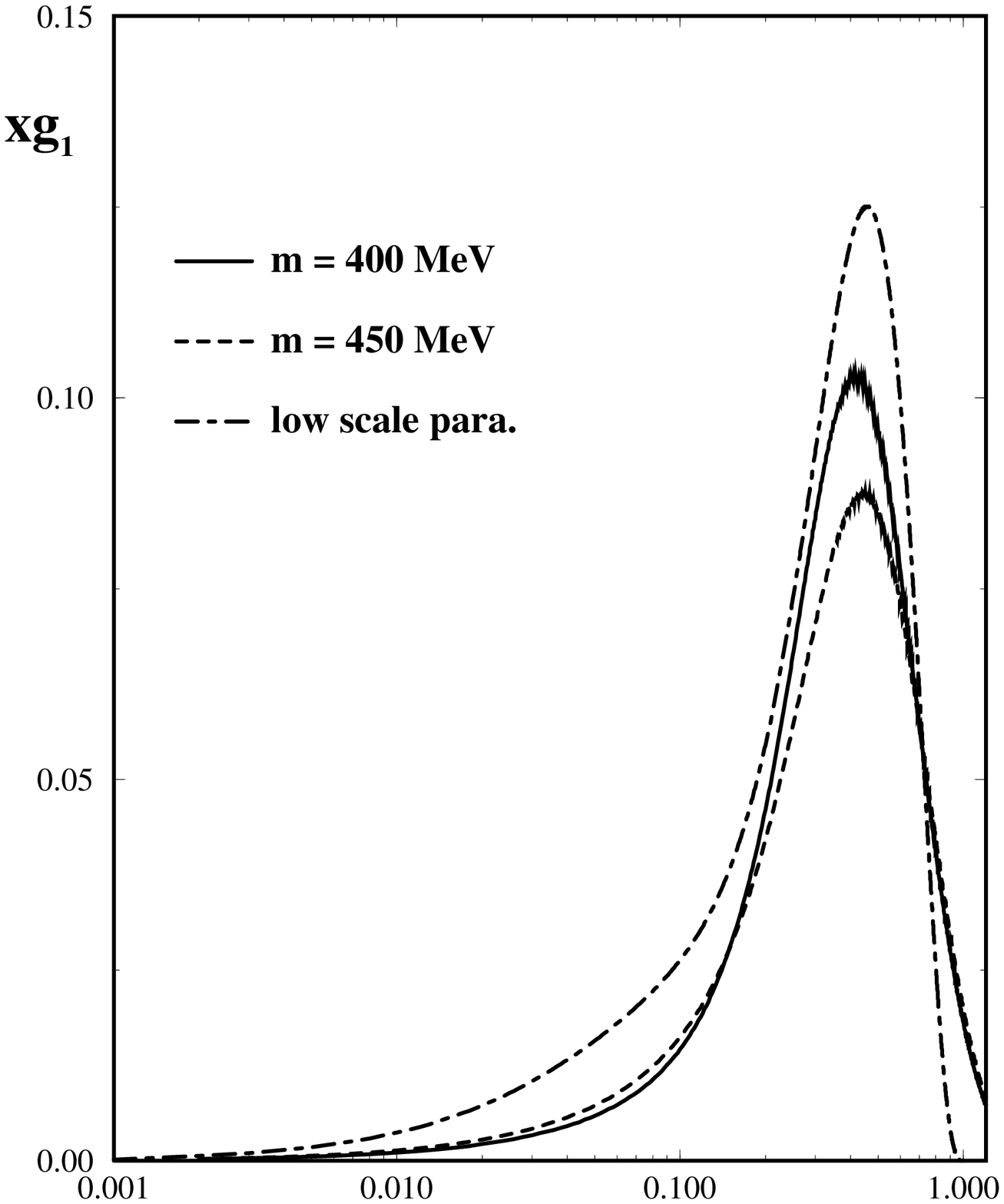,height=5.6cm,width=7.0cm}}
\caption{\label{fig_2}Model structure functions vs.
the low--scale parametrization of ref \protect\cite{Gl95}.
Left panel: The structure function $F_2(x)$ for 
$eN$ scattering. Right panel: The polarized nucleon structure
function $xg_1$.} 
\vspace{0.3cm}
\centerline{\hspace{0.3cm}
\epsfig{figure=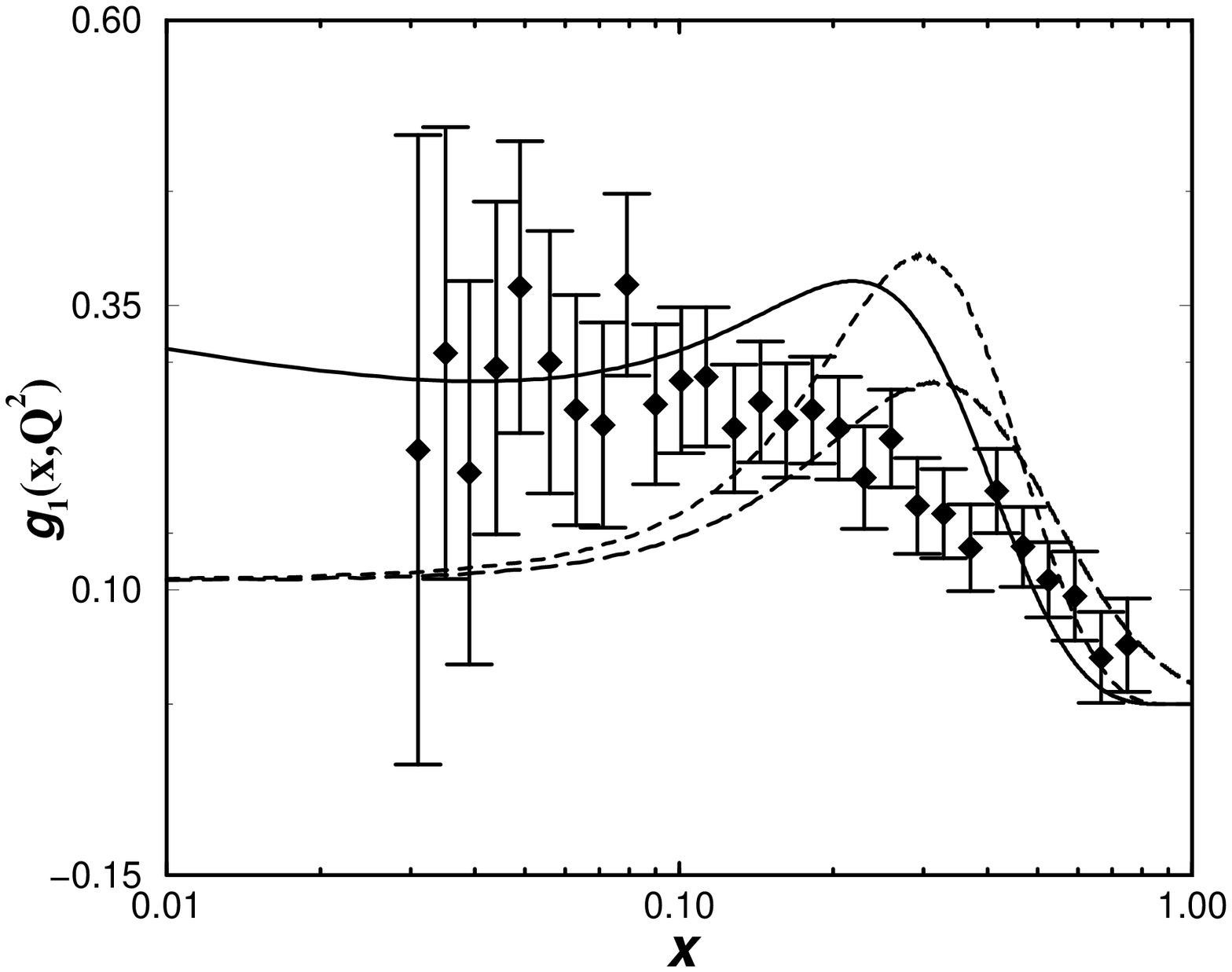,height=6.8cm,width=7.5cm}
\hspace{0.1cm}
\epsfig{figure=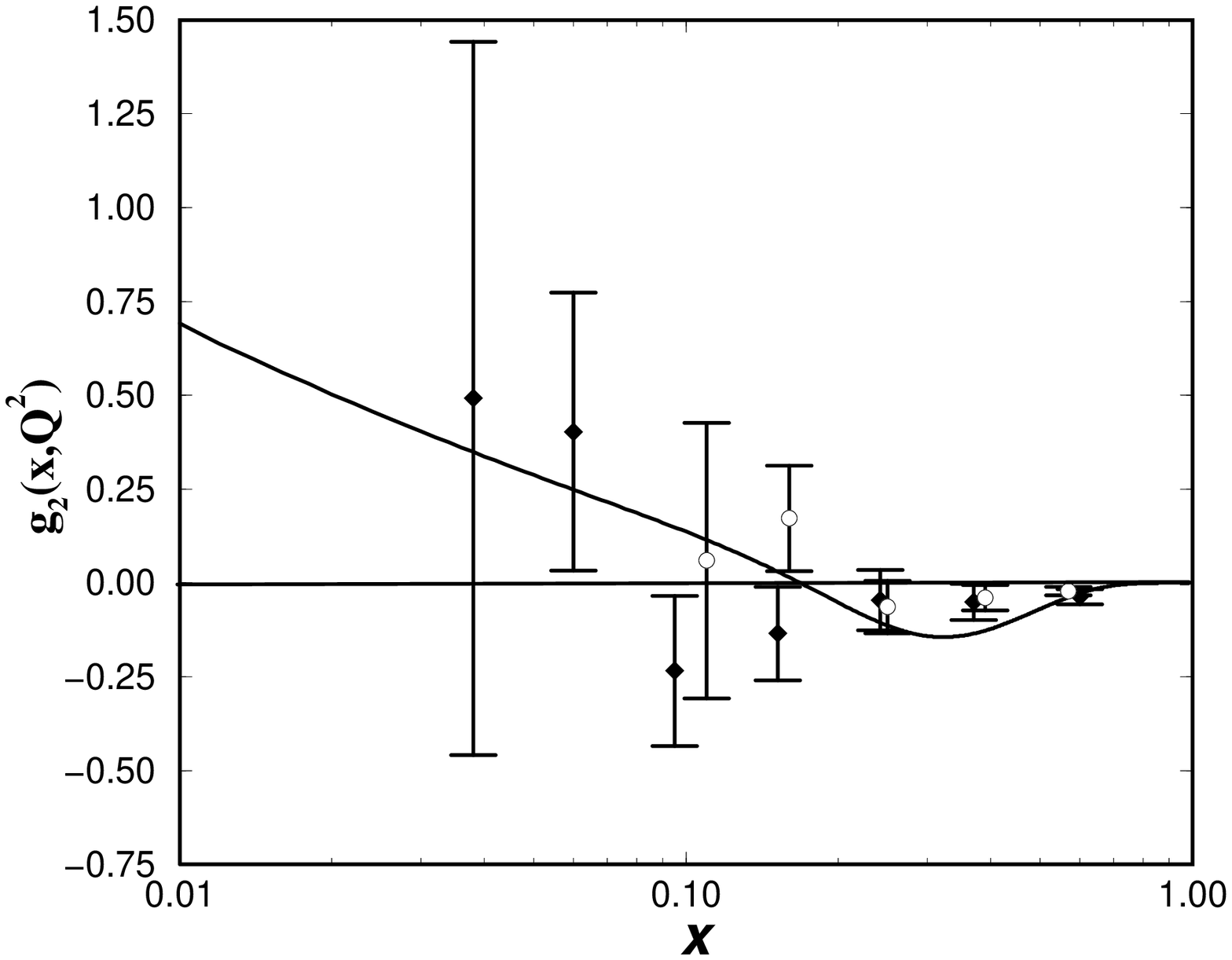,height=6.8cm,width=7.5cm}}
~
\vspace{-1.3cm}
\caption{\label{fig_3}The polarized structure functions $g_1$
and $g_2$ after projection and QCD evolution. Left panel: 
The dashed (dotted) line denotes the (projected) low scale 
model prediction.}
\vspace{-0.4cm}
\end{figure}

In figure \ref{fig_2} 
we display the linear combination relevant for the Gottfried sum rule
\be
\left(F_2^{ep}-F_2^{en}\right)=-
x\left(f_+^1-f_-^1\right)/3
\label{gott}
\ee
and compare it to the low--scale parametrization of the empirical
data \cite{Gl95}. This is obtained from a next--to--leading
order QCD evolution of the experimental to a low--energy regime 
typical for soliton models. 
The model reproduces the gross features of the 
low--scale parametrization. Moreover the integral of the 
Gottfried sum rule
\be
S_G=
\int_0^\infty \frac{dx}{x}
\left(F_2^{ep}-F_2^{en}\right) 
=\cases{0.29\ , \ m=400 {\rm MeV}\cr
0.27\ , \ m=450 {\rm MeV}}
\label{gottrule}
\ee
reasonably accounts for the empirical value $S_G=0.235\pm0.026$
\cite{Ar94}. In particular the deviation from the na{\"\i}ve value 
(1/3) \cite{Go67} is in the direction demanded by experiment. 
In figure \ref{fig_2} we also compare the model prediction for the 
polarized structure function $g_1(x)$ to its low--scale 
parametrization\cite{Gl95}. Apparently they agree the better the 
smaller the constituent quark mass. This behavior is opposite to 
the unpolarized case. No low--scale parametrization is available for 
the polarized structure function $g_2(x)$. Therefore we have 
projected the corresponding prediction onto the interval $x\in[0,1]$ 
\cite{Ja81,Ga97} and subsequently performed a leading order QCD 
evolution to the scale of the experiment\cite{We97b}. Figure 
\ref{fig_3} shows that the resulting polarized structure functions
reproduce the empirical data quite well, although the latter have 
sizable errors.

Note that
for consistency with the Adler sum rule also the moment of inertia
must be restricted to the valence quark contribution
\cite{We96,We97a}. For $m=350{\rm MeV}$ this fortunately is
almost 90\%. Analogously the Bjorken, Burkhardt--Cottingham
as well as the axial singlet charge sum rules are confirmed
within this model treatment\cite{We97b}. 

\section*{Conclusions}

We have presented a calculation of nucleon structure functions 
within a chiral soliton model. We have 
argued that the soliton approach to the bosonized version of 
the NJL--model is most suitable since (formally) the 
current operator is identical to the one in a free Dirac 
theory. Hence there is no need to approximate the current 
operator by {\it e.g.} performing a gradient expansion.
Although the calculation contains a few (well--motivated)
approximations it reproduces the gross features of the empirical
structure functions at low energy scales. This is true
for the polarized as well as the unpolarized structure functions. 
Future projects will include the extension of the valence 
quark approximation, improvements on the projection issue 
and the extension to flavor SU(3).

\bibliographystyle{unsrt}

\end{document}